\documentclass[referee]{mn2e}

\usepackage{graphicx}
\usepackage{amsmath}

\title[Hadronic model for AR Sco]
{Hadronic model for the non-thermal radiation from the binary system AR~Scorpii}
\author[W. Bednarek]
{W. Bednarek \\ 
Department of Astrophysics, The University of \L \'od\'z,
ul. Pomorska 149/153, 90-236 \L \'od\'z, Poland,\\
bednar@uni.lodz.pl}
\begin{document}

\date{Accepted . Received ; in original form }

\pagerange{\pageref{firstpage}--\pageref{lastpage}} \pubyear{2015}

\maketitle

\label{firstpage}
\begin{abstract}
AR Scorpii is a close binary system containing a rotation powered white dwarf and a low mass M type companion star. This system shows non-thermal emission extending up to the X-ray energy range. We consider hybrid (lepto-hadronic) and pure hadronic models for the high energy non-thermal processes in this binary system. Relativistic electrons and hadrons are assumed to be accelerated in a strongly magnetised, turbulent region formed in collision of a rotating white dwarf magnetosphere and a magnetosphere/dense atmosphere of the M dwarf star. We propose that
the non-thermal X-ray emission is produced either by the primary electrons or the secondary $e^\pm$ pairs from decay of charged pions created in collisions of hadrons with the companion star atmosphere. We show that the accompanying $\gamma$-ray emission from decay of neutral pions, that are produced by these same protons, is expected to be on the detectability level of the present 
and/or the future satellite and Cherenkov telescopes.  The $\gamma$-ray observations of the binary system AR Sco should allow to constrain the efficiency of hadron and electron acceleration and also the details of the radiation processes.
\end{abstract}

\section{Introduction}

The pulsed emission (from radio up to ultraviolet) has been recently discovered from the binary system AR Sco
containing a white dwarf (WD) and a low mass M type star (Marsh et al.~2016, Buckley et al.~2017). 
This emission varies with the orbital period of the binary system 3.56 hr and also with the rotation period of the white dwarf 1.97 min. Moreover, the emission is accompanied by the non-thermal X-ray emission in the energy range between 0.3-10 keV (Marsh et al.~2016). X-ray emission shows modulation with a pulse fraction of 14$\%$ (Takata et al.~2017). The spin-period seems to slow down indicating that the white dwarf acts like a pulsar
with the spin down luminosity $\simeq 3\times 10^{33}$ erg s$^{-1}$. 
This is clear evidence that not only neutron stars but also magnetized white dwarfs can act under the pulsar mechanism emitting non-thermal radiation (e.g. Paczy\'nski 1990, Usov~1993, Zhang \& Gil~2005, Ikhsanov \& Biermann~2006, Malheiro et al.~2012, Bednarek~2012). Such WD pulsars are expected to originate in the initial spin-up phase due to the accretion of matter from the companion star (e.g. Beskrovnaya \& Ikhsanov~2016). 
The inferred spin-down luminosity is sufficient to power the observed multi-wavelength emission from the system. The lack of evidence of any mass transfer from the companion star to the white dwarf (e.g. Marsh et al. 2016) suggests that the total spin down reservoir is channelled into non-thermal emission (e.g. Marsh et al. 2016, Buckley et al. 2017), which makes AR Sco unique amongst the close binaries.
Most of the emission observed from AR Sco is interpreted in terms of the synchrotron process of relativistic electrons accelerated in the interaction of the white dwarf magnetosphere with the wind from the M dwarf star (Geng et al.~2016, Buckley et al.~2017). In this model, the white dwarf pulsar injects a magnetized plasma which interact with the stellar wind producing a bow shock. On the other hand, Katz (2017) proposes that the spin down power is dissipated in the direct interaction of the WD magnetic field with a dense matter of the M dwarf star atmosphere. Then, particles can be accelerated in a turbulent collision region either in the Fermi acceleration mechanism and/or the magnetic reconnection. In this paper, we assume that
electrons and/or protons can be accelerated to TeV energies close to the surface of the M star in the turbulent magnetized collision region as postulated by the models mentioned above. 
Relativistic protons interact with the matter in the M dwarf atmosphere ($p + p\rightarrow \pi^\pm + \pi^{\rm 0}$), producing secondary $e^\pm$ pairs and high energy $\gamma$-rays via decay of pions. Primary and/or secondary leptons are responsible for the non-thermal synchrotron X-ray emission from AR Sco. The level of this non-thermal emission allows us to predict the $\gamma$-ray flux that is instantaneously produced by protons.

\section{Acceleration of particles}

\begin{figure}
\vskip 6.2truecm
\includegraphics{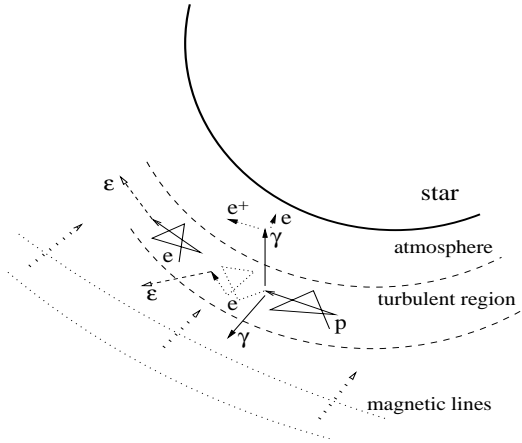}
\caption{Schematic presentation of the considered scenario. The white dwarf magnetosphere collides with the atmosphere of the dwarf M type star forming a turbulent collision region. Protons and electrons can be accelerated in such a turbulent region in the Fermi acceleration mechanism. Relativistic primary electrons cool on the synchrotron process in the magnetic field. Relativistic protons interact with the matter in the atmosphere creating the high energy $\gamma$-rays (from decay of neutral pions) and $e^\pm$ pairs (from decay of charged pions). The
$e^\pm$ pairs cool rapidly close to the production place in the synchrotron process. $\gamma$-rays partially escape and partially move towards denser regions in the atmosphere interacting with the matter ($\gamma + p\rightarrow e^\pm$).}
\label{fig1}
\end{figure}

We assume that particles are accelerated in a turbulent region or a shock formed as a result of the interaction between the WD magnetosphere or the plasma from the white dwarf with the atmosphere (magnetosphere) of the companion star (see schematic representation in Fig.~1).
In such turbulent magnetized plasma, electrons and hadrons can be accelerated in the second order Fermi acceleration mechanism, on small scale shocks (the first order Fermi mechanism) and possibly also in the magnetic reconnection of the turbulent magnetic field which enter the collision region from the site of the WD and the companion star. We consider the process of acceleration of particles in the turbulent plasma by parametrising their acceleration time scale by,
\begin{eqnarray}
\tau_{\rm acc} = R_{\rm L}/c\chi\approx  1E_{\rm TeV}/\chi_{-3}B_2~~~{\rm s}, 
\label{eq1}
\end{eqnarray}
\noindent
where $R_{\rm L}$ is the Larmor radius of the particle, $E_{\rm e,p} = 1 E_{\rm TeV}$ TeV is the particle energy, $\chi = 10^{-3}\chi_{-3}$ is the so called acceleration coefficient, $B = 100B_2$ G is the magnetic field strength in the acceleration region, and $c$ is the speed of light. The value of the acceleration coefficient, $\chi$, is related to the characteristic velocities of the scattering centres
in the turbulent plasma. It can be estimated from $\chi\sim 0.1\beta^2\sim 10^{-3}$ (Malkov \& Durry~2001), where $\beta = v/c$ is the relative velocity of the scattering centres in respect to the velocity of light $c$. $v$ is assumed to be smaller but of the order of the rotation velocity, $v_{\rm rot}$, of the WD magnetosphere at the location of the companion star. $v_{\rm rot}$ is equal to $2\pi d/P_{\rm WD}\approx 4.1\times 10^9$ cm~s$^{-1}$, where the rotation period of the WD is $P_{\rm WD} = 117$ s and the distance between the companion stars is $d\sim 7.6\times 10^{10}$ cm (Marsh et al.~2016). Such a simple description of the acceleration process is frequently used in modelling of the radiation processes in cosmic sources due to still not well known details of the physics of complicated processes related to the injection, acceleration, and interaction of particles 
with the magnetized medium (e.g. see review article by Malkov \& Durry~2001).

The acceleration process of electrons is usually saturated by their synchrotron energy losses.
The cooling time scale of electrons in the synchrotron process is,
\begin{eqnarray}
\tau_{\rm syn} = 3m_{\rm e}c^2/4c\sigma_{\rm T}U_{\rm B}\gamma_{\rm e}\approx  2.8\times 10^{-2}/(E_{\rm TeV}B_2^2)~~~{\rm s}, 
\label{eq2}
\end{eqnarray}
\noindent
where $\gamma_{\rm e} = E_{\rm e}/m_{\rm e}c^2$ is the Lorentz factor of electrons, $U_{\rm B} = B^2/8\pi$ is the energy density of the magnetic field, $\sigma_{\rm T}$ is the Thomson cross section, 
and $m_{\rm e}$ is the electron rest mass. The magnetic fields of the order of a hundred Gauss are expected to be provided to the turbulent region
from the WD magnetosphere (e.g. Katz~2017, Takata et al. 2017). On the other hand, the polar surface magnetic fields of the low mass companion stars can be up to $\sim 3000$ G (e.g. Meintjes \& Jurua~2006, Donati \& Landstreet~2009).
Therefore, these turbulent regions are expected to be strongly magnetized allowing for very efficient cooling of
accelerated electrons.

By comparing these two time scales, we estimate the characteristic energies of accelerated electrons,
\begin{eqnarray}
E_{\rm e}^{\rm max}\approx 0.17(\chi_{-3}/B_2)^{1/2}~~~{\rm TeV}. 
\label{eq3}
\end{eqnarray}
\noindent
Then, the maximum energies of synchrotron photons produced by these electrons are,
\begin{eqnarray}
\varepsilon_{\rm syn}\approx m_{\rm e}c^2(B/B_{\rm cr})\gamma_{e^\pm}^2\sim 100\chi_{-3}~~~{\rm keV}, 
\label{eq4}
\end{eqnarray}
\noindent
where $B_{\rm cr} = 2\pi m_{\rm e}^2c^3/he = 4.4\times 10^{13}$ G is the critical magnetic field strength,
$m_{\rm e}$ is the electron rest mass, $h$ is the Planck constant, and $e$ is the charge of electron.
We have calculated the synchrotron spectrum produced in the process of a complete cooling of electrons with the power law spectrum and an exponential cut-off at $E_{\rm e}^{\rm max}$, i.e. $dN_{\rm e}/dE_{\rm e}\propto E_{\rm e}^{-\alpha}\exp(-E_{\rm e}/E_{\rm e}^{\rm max})$, where $\alpha$ is the spectral index. The {\it Swift} X-ray spectrum is reported to have the spectral index close to -2 in the energy range 0.3$-$10 keV (Marsh et al.~2016). More recent analysis of the Swift data reports the spectral index of the pulsed component 
equal to 2.3$\pm$0.5 (Takata et al. 2017).
The synchrotron spectrum with such spectral index is produced by a completely cooled electrons that are injected also with the spectral index close to -2. In fact, electrons with such spectral index are expected to be accelerated in the Fermi process.

We also consider the possibility that the non-thermal X-ray emission is produced in the synchrotron process by 
leptons which are secondary products of the interactions of relativistic protons with the matter in the companion star atmosphere. Protons are accelerated in the mentioned above turbulent region of the WD magnetosphere. The acceleration of protons is limited by their escape from the turbulent region. The escape time scale is estimated as,   
\begin{eqnarray}
\tau_{\rm esc} = 2R_\star/v_{\rm adv}\approx  50/v_{9}~~~{\rm s}, 
\label{eq5}
\end{eqnarray}
\noindent
where $v_{\rm adv} = 10^9v_{9}$~cm~s$^{-1}$ is the characteristic advection velocity assumed to be 
smaller than the velocity of the magnetic field lines of the WD at the location of the companion star, 
and $R_\star = 2.5\times 10^{10}$ cm is the radius of the star.
By comparing the acceleration time scale of protons with their escape time scale, we obtain the maximum energies
of accelerated protons,
\begin{eqnarray}
E_{\rm p}\approx  50B_2\chi_{-3}/v_{9}~~~{\rm TeV}. 
\label{eq6}
\end{eqnarray}
For the magnetic field strengths mentioned above, i.e. $\sim 100$ G, the acceleration coefficient 
$\xi = 10^{-3}$ and the advection velocity $v_{\rm adv} = 10^9$ cm~s$^{-1}$, protons are expected to be accelerated up to $\sim 100$ TeV. Note that the Larmor radius of protons with such energies in the magnetic field of 100 G, equal to $R_{\rm L} = 3\times 10^9$ cm, is about an order of magnitude smaller than the radius of the companion star.

Hadrons with the multi-TeV energies can find enough target in the atmosphere of the dwarf star suffering efficient hadronic collisions with the matter. They are expected to produce 
$\gamma$-rays and $e^\pm$ pairs from decay of pions ($p + p\rightarrow \pi^{\rm 0} + \pi^\pm$). 
We assume that the non-thermal X-ray emission, observed from the binary system AR Sco, can be produced by those secondary $e^\pm$ pairs in the synchrotron process. The energies of these $e^\pm$ pairs 
can be simply estimated from,
$E_{e^\pm}\approx \kappa E_{\rm p}/4\mu_{pp}\sim 1$~TeV, $\kappa\approx 0.5$ is the inelasticity in the p-p interactions, $\mu_{pp}\approx 2.57\times \ln(2E_{\rm GeV}) - 6.45$ is the pion multiplicity in p-p interactions, and $E_{\rm GeV}$ is the proton energy in GeV (Orth \& Buffington~1976).
The $e^\pm$ pairs produce characteristic synchrotron radiation with energies,
\begin{eqnarray}
\varepsilon_{\rm syn}\approx m_{\rm e}c^2(B/B_{\rm cr})\gamma_{e^\pm}^2\approx 5B_2~~~{\rm MeV}. 
\label{eq7}
\end{eqnarray}
\noindent
Therefore, hadronic model predicts the synchrotron spectrum from secondary $e^\pm$ pairs extending through
the hard X-ray up to the soft $\gamma$-ray energy range. Since the cooling time scale of $e^\pm$ pairs is short (see estimates above), these pairs lose energy already in the turbulent region. However, the fate of the $\gamma$-rays, produced from decay of neutral pions, can be different. A significant part of these $\gamma$-rays is expected to be directed towards the dense atmosphere of the companion star due to the geometrical effects and also due to the collimation of charged protons by the ordered component of the magnetic field of the companion star. Therefore, $\gamma$-rays can be partially absorbed in the inner dense atmosphere of the companion star since the cross section for their absorption in the matter ($\gamma + p\rightarrow e^\pm$) is of the order of the cross section for the pion production in collisions of relativistic protons with the matter (e.g. Lang~1999). 

Below we calculate the synchrotron spectra produced by the secondary $e^\pm$ pairs and also the $\gamma$-ray spectra from decay of pions. From the fitting of the observed X-ray spectrum by the synchrotron emission from the secondary $e^\pm$ pairs, we determine the power which should be transferred to the relativistic  
electrons. Based on this normalization, we predict the expected $\gamma$-ray spectrum in the GeV-TeV energy range in terms of such hadronic model. We show that the hadronic model can be tested by the observations with the HESS array, the planned Cherenkov Telescope Array (CTA) and also through the analysis of the available ${\it Fermi}$-LAT data.

\section{Non-thermal radiation}

The non-thermal X-ray emission from AR Sco indicates on the presence of relativistic
electrons in this source. However, it is not clear whether electrons are accelerated directly in the turbulent collision region between the WD magnetosphere and the companion star atmosphere or they are the secondary products from other processes. Therefore, we consider two general scenarios. In the first one, electrons and protons are accelerated with a specific ratio of the powers described by the factor $\eta = L_{\rm e^-}/L_{\rm p}$ and the spectral indexes equal to -2. In this case, we interpret the {\it Swift} X-ray observations of AR Sco as due to the synchrotron emission from the primary electrons in the strong magnetic field of the turbulent region (see Fig.~2a). The obtained fit is consistent with the lower limit on the Swift X-ray spectrum. The upper limit will require flatter injection spectrum of particles than -2.  In principle, it can be obtained in the acceleration process in the magnetic reconnection regions. However, the fitting of the Swift upper limit on the X-ray flux does not have significant effect on the predicted $\gamma$-ray spectrum from decay of pions since the Swift fit is obtained by the upper energy part of the synchrotron spectrum.
The power transferred to relativistic electrons has to be $2.6\times 10^{31}$ erg~s$^{-1}$ in order to explain the observed level of the {\it Swift} emission and assuming the complete cooling of electrons. This means that the WD should transfer about $1\%$ of its rotation energy to the relativistic electrons. 
In this scenario we postulate that the acceleration and radiation processes of primary electrons occur at a relatively rare parts of the companion star atmosphere. However, even if it occurs in a quite dense region of the companion star atmosphere, then the synchrotron process of the primary electrons likely overcomes their possible cooling on the bremsstrahlung process since $\tau_{\rm brem} = 1.3\times 10^{15}/n~[s]\gg \tau_{\rm syn}$, where $n$ is the density of matter in $cm^{-3}$ (see Eq.~2 and Lang~1999). 
Electrons with sub-TeV energies are captured effectively by the strong magnetic field since their Larmor radii
are much lower than the radius of the companion star. Therefore, X-rays, propagating along the straight lines,  are expected to escape without significant absorption in the matter of the WD atmosphere.

However, relativistic protons are also likely to be accelerated in the turbulent region.
Their acceleration site should be linked with the dense regions of the M dwarf companion star by its large scale
magnetic field. Therefore, protons are expected to be driven towards dense regions of the stellar atmosphere. They can efficiently interact with the matter of the companion star atmosphere producing secondary $e^\pm$ pairs and $\gamma$-rays from decay of pions. We calculate the synchrotron emission from the secondary 
$e^\pm$ pairs and the accompanying $\gamma$-ray emission assuming that these particles are produced in the pion decay originated in a single interaction of protons (see Fig.~2a) and in the process of complete cooling of protons (Fig.~2b). If more energy is transferred to relativistic protons than to electrons, then the synchrotron spectra (from secondary $e^\pm$ pairs) and the $\gamma$-ray spectra are proportionally enhanced (see the dotted curve for the ratio of the electron to proton powers equal to $\eta = 1/3$ and the dot-dashed curve for $\eta = 1/10$ in Figs.~2a,b). 

\begin{figure*}
\vskip 5.truecm
\includegraphics{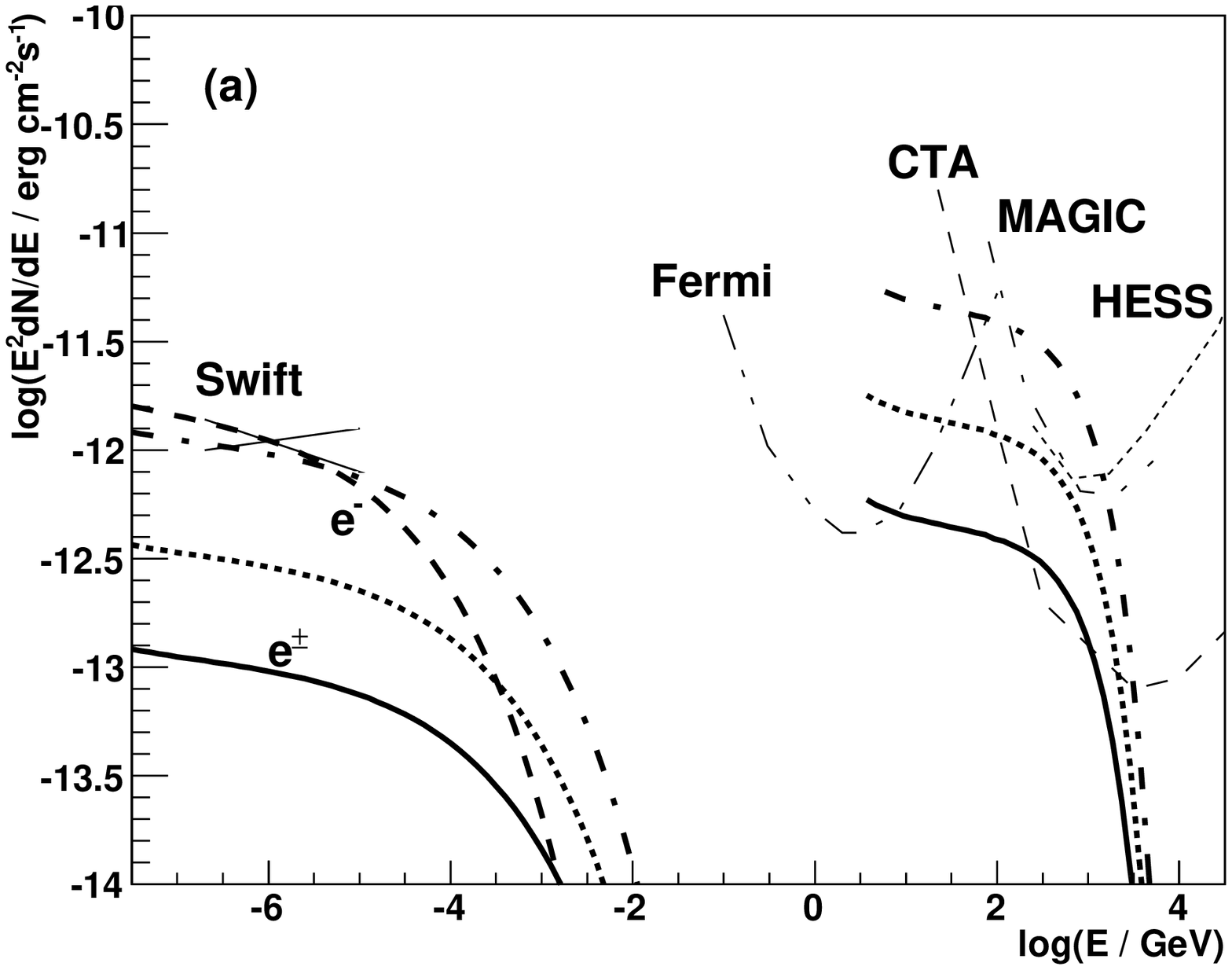}
\includegraphics{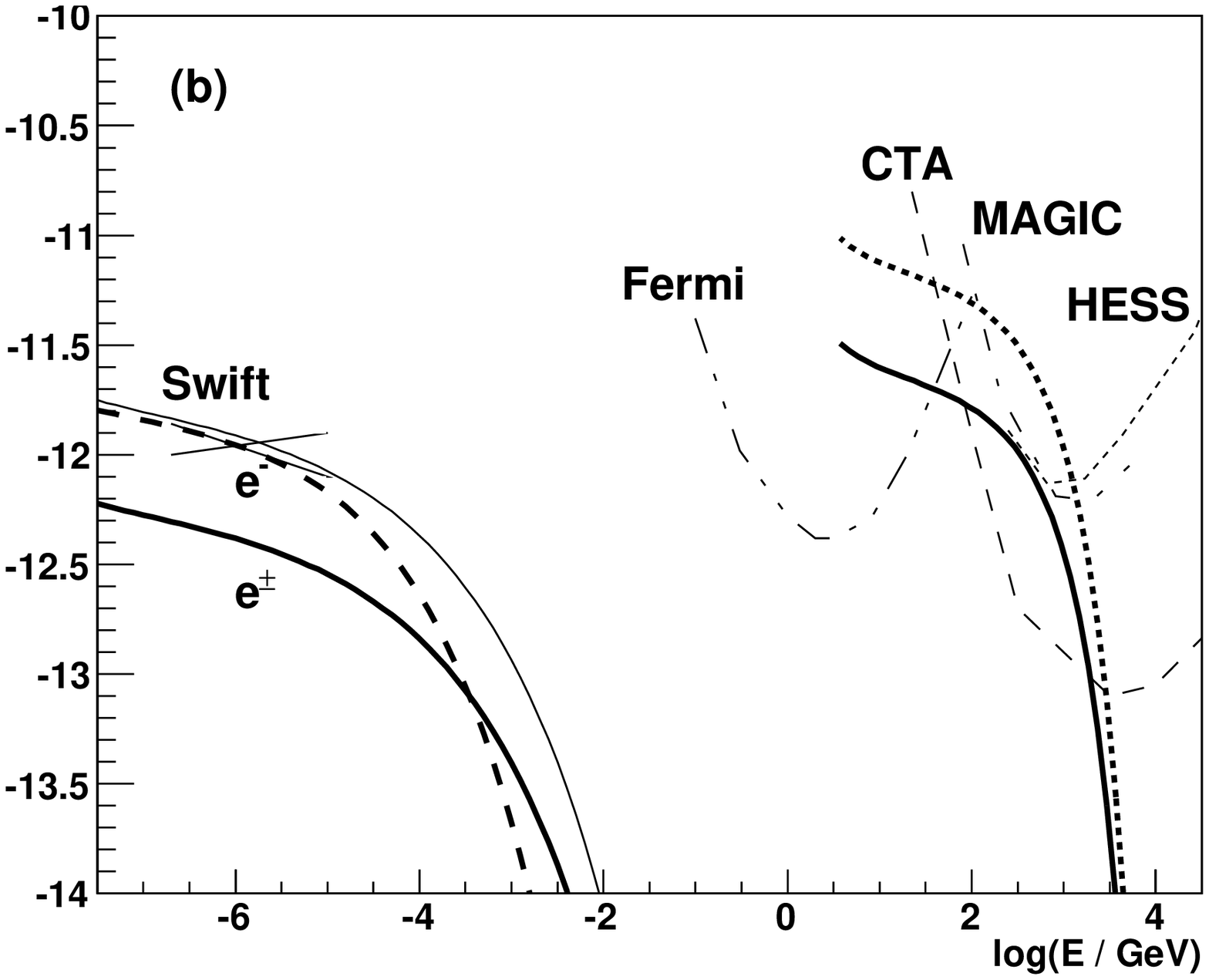}
\includegraphics{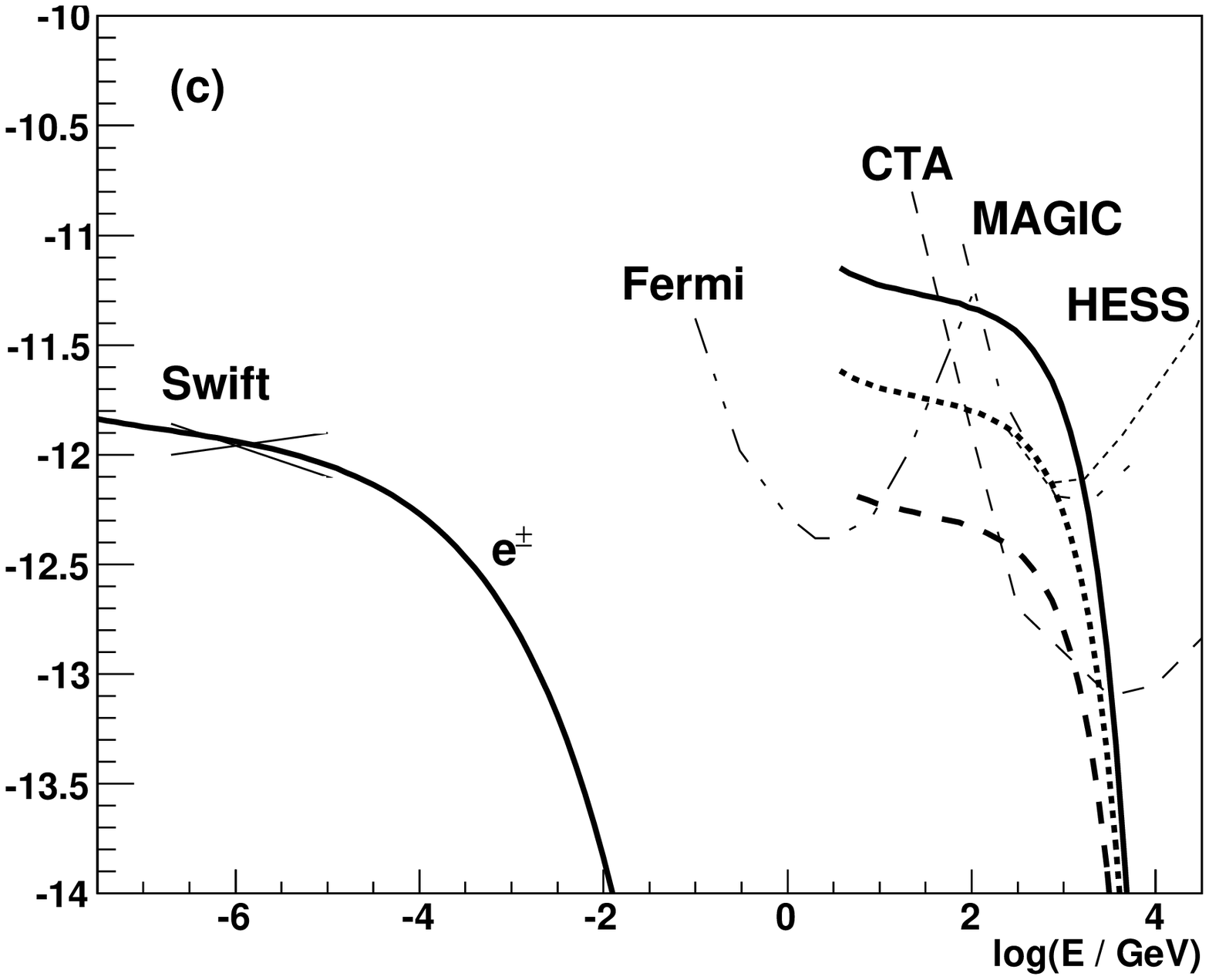}
\caption{(a) The non-thermal X-ray spectrum observed by the {\it Swift} satellite from the binary system AR Sco (Marsh et al.~2016, thin solid lines) is compared with the synchrotron emission of the primary electrons ($e^-$, dashed curves) and the synchrotron emission of the secondary $e^\pm$ pairs ($e^\pm$, solid curve). The secondary pairs are produced in the decay process of charged pions originated in a single collision of relativistic protons with the matter in the M dwarf companion star atmosphere. It is assumed that the electrons and protons obtain the power law spectra with the spectral index close to -2 with the exponential cut-off at some maximum energies defined in Sect.~2. The $\gamma$-ray spectrum from decay of neutral pions, produced by the relativistic protons with the power equal to the power of relativistic electrons, i.e. $\eta = L_{\rm e^-}/L_{\rm p} = 1$, is shown by the solid curve. The synchrotron and $\gamma$-ray emission in the case of $\eta = 1/3$ and 1/10 are marked by the dotted and dot-dashed curves, respectively. In order to fit the X-ray emission, the primary electrons has to 
gain  $L_{\rm e^-} = 2.6\times 10^{31}$ erg~s$^{-1}$ energy from the acceleration mechanism (see dashed curve marked by $e^-$). The synchrotron emission produced by the secondary $e^\pm$ pairs from decay of charged pions ($p + p\rightarrow \pi^\pm$) and $\gamma$-rays from decay of neutral pions ($p + p\rightarrow \pi^{\rm 0}$) are marked by the solid curves. (b) As in figure (a) but for the case of a complete cooling of relativistic protons in the interaction with the matter. (c) In this model only protons are accelerated. The {\it Swift} X-ray emission is described by the synchrotron emission from the secondary $e^\pm$ pairs from decay of charged pions. Pions (and $\gamma$-rays and $e^\pm$ pairs) are produced in a single interaction of protons with the matter (solid curve). We accept that the $\gamma$-rays can be also partially absorbed in the matter assuming that only 30$\%$ of $\gamma$-rays (dotted curve) or 10$\%$ of $\gamma$-rays (dashed curve) are able to escape without absorption. The $\gamma$-ray emission is confronted with the sensitivities of the {\it Fermi}-LAT (10 yrs, see dot-dot-dashed curve, Funk et al. 2013), the HESS array (25 hrs, thin dotted curve, HESS Colab.) the MAGIC array (50 hrs, thin dot-dashed curve, Aleksi\'c et al. 2012) and the planned CTA array (50 hrs, thin dashed curve, Maier et al.~2017).}
\label{fig2}
\end{figure*}

We also consider the non-thermal processes in AR Sco under the hypothesis that the {\it Swift} X-ray emission is produced only by the secondary $e^\pm$ pairs, i.e. primary electrons are not accelerated to large energies in the turbulent region (see Fig.~2c).
In this case we also show the $\gamma$-ray spectra that might be significantly absorbed in collisions with the matter of the atmosphere in the process $\gamma + p\rightarrow e^\pm$ (see dashed and dotted curves in Fig.~2c) for the case in which only $10\%$ and $30\%$ $\gamma$-rays escape from the companion star atmosphere).

It is expected that the column density of the matter traversed by relativistic protons is much larger than that traversed by $\gamma$-rays. In fact, protons will be captured by the magnetic field on a characteristic distance scale corresponding to their Larmor radii, $R_{\rm L} = E/eB\approx 3\times 10^7 E_{\rm TeV}/B_2$~cm. Then, the column density traversed by rectilinearly propagating $\gamma$-rays should be a factor of $R_{\rm L}/a << 1$ lower. Therefore, it is expected that protons can efficiently produce $\gamma$-rays in collisions with the matter of the companion star atmosphere. On the other hand, produced $\gamma$-rays can escape without significant absorption due to the relatively low column density of matter along their straight paths. Due to the effective capturing of relativistic protons, the production of secondary leptons can already occur in a relatively rare matter allowing free escape of produced by them synchrotron X-rays without strong absorption.

Note that in the optimistic case, i.e. more energy transferred to relativistic protons than to relativistic electrons ($\eta \ll 1$) and the level of $\gamma$-ray absorption is low, the $\gamma$-ray fluxes are predicted within the sensitivity limits of the present Cherenkov telescopes in the TeV energy range. In fact, the HESS array have the best conditions for the observations of AR Sco since it is located on the southern hemisphere. 
The spectrum predicted in the optimistic case should be detected by the HESS in 2-3 hours.
This $\gamma$-ray emission should be also detected in the GeV energy range by the ${\it Fermi}$-LAT telescope
in about a month. However, in the pessimistic case, i.e. the ratio of $\eta\sim 1$ (Fig.2a) or only $10\%$ of $\gamma$-rays escape from the companion atmosphere (Fig.~2c), the $\gamma$-ray emission is close to the 30 hours sensitivity of the planned CTA and also within 8 years of sensitivity of the {\it Fermi}-LAT telescope. 
Therefore, we conclude that the extensive observations of the binary system AR Sco at $\gamma$-ray energies can provide important constraints on the acceleration processes of different types of particles (leptons and hadrons) in the magnetised turbulent plasma formed in the collision of the rotating magnetosphere of the WD with the dense atmosphere of a companion star.

In the above calculations we neglected the energy losses of relativistic electrons in the inverse Compton scattering (ICS)
of thermal radiation emitted from the surface of the M star and the white dwarf in respect to their energy losses on the synchrotron process. These energy losses depend on the energy densities of the magnetic and
radiation fields in the acceleration region of electrons.
The basic parameters of these stars has been estimated on: surface temperature of M star $T_\star = 3100$~K and its radius $2.5\times 10^{10}$~cm and the white dwarf $T_\star = 9750$~K and the radius $7\times 10^{8}$~cm (Marsh et al. 2016).
Then, the energy density of radiation from the M star is estimated on $\sim$0.7~erg~cm$^{-3}$ and from the white dwarf on
$\sim 70(R_{\rm WD}/a)^2\sim$0.07~erg~cm$^{-3}$. In the case of the WD we take into account that its radiation
is diluted due to the large separation of the stars within the binary system $a = 8\times 10^{10}$~cm. On the other hand, the energy density of the magnetic field with the applied strength of 100 Gs is $\sim 440$~erg~cm$^{-3}$.
We conclude that the IC energy losses of electrons can be safely neglected in respect to their synchrotron energy losses.
The IC $\gamma$-ray spectra produced by electrons should be by a factor of $\sim 10^{-3}$ lower than the synchrotron spectra shown in Fig.~2.

\section{Conclusion}

We consider the consequences of the hypothesis that both electrons and protons are being accelerated within the binary system AR Sco. Protons can be accelerated to the multi-TeV energies close to the dense atmosphere of the M type star. Thus, they can find enough target for efficient interaction. As a result, they produce secondary $e^\pm$ pairs and $\gamma$-rays from decay of pions. We interpret the observed non-thermal X-ray emission from this binary in terms of three scenarios. In the first one, both electrons and protons are accelerated with some ratio of the powers $\eta$. The synchrotron emission of electrons is responsible for the observed non-thermal X -ray emission. On the other hand, $\gamma$-rays originate from the decay of neutral pions that are in turn produced by protons in a single interaction with the matter of the stellar atmosphere.
The $\gamma$-ray emission can be detected by the {\it Fermi}-LAT and future CTA provided that $\eta\sim 1$ (see Fig.~2a). The present Cherenkov telescopes (e.g. the HESS array) could detect this $\gamma$-ray emission provided that $\eta < 1/3$. In the second scenario, the X-ray emission is explained by the synchrotron emission of primary electrons (and/or secondary $e^\pm$ pairs) but protons are transported to dense regions of the M dwarf atmosphere suffering complete cooling in the hadronic collisions. Then, the $\gamma$-ray fluxes are clearly larger (see Fig.~2b). These $\gamma$-rays have a chance to be detected by the HESS array provided that $\eta < 1$ and by the {\it Fermi}-LAT for $\eta < 3$, i.e. the power in relativistic electrons is a factor of 3 larger than the power in relativistic protons. In the third scenario, we assumed that only protons are efficiently accelerated. Then, the non-thermal X-ray emission is explained by the synchrotron radiation from the secondary $e^\pm$ pairs. The $\gamma$-ray emission is expected to be clearly above the sensitivity limits of the {\it Fermi}-LAT telescope and the HESS array (see Fig.~2c). However, this emission might be additionally influenced by the absorption of $\gamma$-rays in the matter ($\gamma + p\rightarrow e^\pm$). We conclude that even if only a small part of the $\gamma$-rays can escape from the M dwarf atmosphere (e.g. $\mu = 10\%$), then their flux should be within the sensitivity of the {\it Fermi}-LAT telescope and the HESS array. A partial absorption of $\gamma$-rays might in principle also affect the constraints obtained above in terms of the first and the second scenarios. Therefore, in principle the limits obtained above should refer to $\eta/\mu$ but not only to $\eta$. The predictions of the GeV-TeV $\gamma$-ray emission from AR Sco, obtained in terms of the hadronic model, look quite encouraging. They could be tested by the present and future Cherenkov telescopes and also by the analysis of the available data collected by the {\it Fermi}-LAT telescope.           

Since AR Sco is the only discovered binary system in which the companion M type star is immersed directly in the inner, rigidly rotating magnetosphere of a compact object (in this case the white dwarf) similar processes are not expected to occur in other type of binary systems with compact objects such as LMXBs (containing accreting neutron stars), black widow/red-back binaries (containing millisecond pulsars) or binary systems of classical radio pulsars with massive companion stars such as 
PSR 1259-63/LS2883. Some of those compact binaries emit (or are expected to emit) high energy $\gamma$-rays in different geometrical and radiative scenarios.

\section*{Acknowledgements}
I would like to thank the Referee for useful comments.
This work is supported by the grant through the Polish Narodowe Centrum Nauki No. 2011/01/B/ST9/00411.


\label{lastpage}
\end{document}